\documentstyle[aps,epsf,prl,twocolumn]{revtex}
\begin{document}
\title{Solid-State Quantum Computer Based on Scanning Tunneling Microscopy}

\author{{G.P. Berman$^1$, G.W. Brown$^2$, M.E. Hawley$^2$, and  V.I. 
Tsifrinovich$^3$}}

\address{$^1$ T-13 and CNLS, 
Los Alamos National Laboratory, Los Alamos, New Mexico 87545}
\address{$^2$MST-8, Los Alamos National Laboratory, MS G755, Los 
Alamos NM 87545}
\address{$^3$IDS Department, Polytechnic University,
Six Metrotech Center, Brooklyn NY 11201}

\maketitle

\begin{abstract}
We propose a solid-state nuclear spin quantum computer based on 
application of scanning tunneling microscopy (STM) and well-developed 
silicon technology. It requires the measurement of tunneling current 
modulation caused by the Larmor precession of a single electron spin.
  Our envisioned STM quantum computer would operate at the high magnetic 
field ($\sim 10$T) and at low temperature $\sim 1$K.
\newline
{PACS numbers: 03.67.Lx,~03.67.-a,~76.60.-k}

%
\end{abstract}
\section{Introduction}
Recently, we suggested a solid-state nuclear spin quantum computer based on
   magnetic resonance force microscopy (MRFM) \cite{1,2}. In this 
proposal, a ferromagnetic particle, which is attached to the tip of 
the cantilever, moves along a chain of paramagnetic ions (atoms) 
located below the surface of
  silicon. The ferromagnetic particle targets a particular ion, 
providing selective excitation of its electron or nuclear spin, a 
measurement of the state of the nuclear spin, and initial polarization 
and one-qubit rotations of nuclear spins. A two-qubit Control-Not 
(CN) gate for two neighboring nuclear spins utilizes the 
dipole-dipole interaction between the nuclear spin of one ion and the 
electron of the neighboring ion.

There are three basic disadvantages in the suggested MRFM proposal: 
1) The motion of the ferromagnetic particle along the chain of 
paramagnetic ions can induce additional dephasing in the spin system. 
This dephasing must be taken into account. 2) Fluctuations of the 
position of a ferromagnetic particle cause additional decoherence in 
a spin system. 3) The accurate positioning of the ferromagnetic 
particle relative to an ion is a complicated experimental problem.

In this paper, we suggest a quantum computer architecture which seems 
to be free of all these disadvantages. At the same time, it retains 
the main advantage of the MRFM proposal -- it does not require an 
array of single-atom gates, which is an essential part of well-known 
solid-state proposals based on silicon technology \cite{3,4}. We give 
estimates for characteristic parameters required for realization of a 
quantum computer based on the STM technique.

\section{STM detection of an electron spin resonance and a nuclear spin state}

Our proposal utilizes recent results in STM experiments which seem to 
allow detection of the precessing electron spin from a single ion 
(atom). In these experiments \cite{5}, the STM technique was used to 
measure the frequency of the Larmor precession of an electron spin of an 
individual iron atoms in silicon in the presence of a small, 
static applied magnetic field. This measurement utilizes the 
interaction between a localized electron spin in a surface of a 
sample and the current produced by tunneling electrons. Due to this 
interaction the tunneling probability is affected by
  the Larmor precession of a localized spin, and the tunneling current 
is modulated at the Larmor frequency.
\begin{figure}
\epsfxsize 4cm
\epsfbox{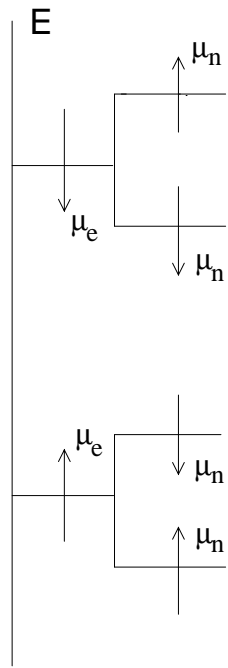}
\caption{Fig. 1:~Energy levels for electron ($\mu_e$) and nuclear 
($\mu_n$) magnetic moments in a high magnetic field.}
\label{fig.1}
\end{figure}
The mechanism causing the tunnel current modulation is, however, not well 
understood. While this requires additional consideration, in this 
paper we assume that the STM technique is able to detect the 
frequency of electron spin precession for an individual ion (atom) in 
a silicon surface. This results in the possibility for the STM 
technique to detect the state of a nuclear spin in the same ion. 
Indeed, suppose that the hyperfine interaction between electron and 
nuclear spins of the ion is greater than the Zeeman interaction 
between a high external magnetic field, $\vec B_0$, and the nuclear 
spin. Then, the frequency of an electron spin precession depends on 
the state of the nuclear spin. As an example, following \cite{2}, we 
consider a $^{125}Te^+$ ion (A-center) in the surface of silicon. The 
Hamiltonian of the electron-nuclear spin system of the ion has the 
form,
$$
{\cal H}=g_e\mu_BB_0S_z+g_n\mu_nB_0I_z-A\vec{S}\vec{I},\eqno(1)
$$
where $\vec{S}$ and $\vec{I}$ are electron and nuclear spin operators 
($S=I=1/2$); $B_0$ is a high magnetic field pointed in the positive 
$z$-direction; $g_e\approx 2$, $g_n\approx 0.882$; and 
$A/2\pi\hbar\approx 3.5$GHz is the constant of the hyperfine 
interaction. The magnetic moments of the $^{125}Te$ nucleus and 
the electron are both negative. Fig. 1, demonstrates the energy levels of 
the electron and nuclear spins of the $^{125}Te^+$
ion in a high magnetic field.

If the nuclear magnetic moment points ``up'', the modulation frequency is,
$$
f=f_{e0}=g_e\mu_BB_0/2\pi\hbar+A/4\pi\hbar.\eqno(2)
$$
In the opposite case,
$$
f=f_{e1}=g_e\mu_BB_0/2\pi\hbar-A/4\pi\hbar.\eqno(3)
$$
The hyperfine splitting is $A/2\pi\hbar$. If the tunneling current is 
observed to oscillate with the frequency $f_{e0}$,
the nuclear spin of the ion is in its ground state. Otherwise the 
nuclear spin is in the excited state.

\section{Selective excitation of\\ a nuclear spin}

To provide a selective excitation of the nuclear spin (qubit), we 
assume that a permanent magnetic field is slightly non-uniform in the 
$x$-direction (see Fig. 2). Assuming that the distance between the 
neighboring ions is $a$, we have, for the difference in the 
precession frequencies for these ions,
$$
\Delta f_e=(g_e\mu_B/2\pi\hbar)a{{\partial B_0}\over{\partial x}}.\eqno(4)
$$
The corresponding difference for nuclear magnetic resonance (NMR) 
frequencies is,
$$
\Delta f_n=(g_n\mu_n/2\pi\hbar)a{{\partial B_0}\over{\partial x}}.\eqno(5)
$$
(In Eq. (4) we assume that nuclear spins of the neighboring ions are 
in the same state. In Eq. (5) we assume that the electron spins of 
the neighboring ions are in the same state.) As an example, taking 
$a=5$nm and $\partial B_0/\partial x=10^5$T/m, we obtain for 
$^{125}$Te: $\Delta f_n\approx 6.75$kHz.

For selective excitation of a nuclear spin, one can apply a $\pi$- or 
$\pi/2$-pulse  of a rotating magnetic field, $\vec B_\perp$. The 
nuclear Rabi frequency, $f_{nR}$, is given by,
$$
f_{nR}=(g_n\mu_n/2\pi\hbar)B_\perp.\eqno(6)
$$
For selective excitation, this quantity must be less than $\Delta 
f_n$. It imposes a restriction on the duration, $\tau$, of a 
$\pi$-pulse,
$$
\tau=\pi/2\pi f_{nR}>1/2\Delta f_n\approx 74\mu s.\eqno(7)
$$
Using selective nuclear $\pi$-pulses one can drive a nuclear spin 
chain into the ground state. For this purpose one prepares the system 
in a high magnetic field at a low temperature, so that all electron 
spins are in their ground state. (As an example, $B_0\sim 10$T, 
$T\sim 1$K.) Then, one measures the modulation frequency of the 
tunneling current at any position, $x_k$. If the modulation frequency 
fits the value, $f_{e1}(x_k)$, one applies a selective nuclear 
$\pi$-pulse with the frequency, $f_n(x_k)$, which drives the nuclear 
spin into the ground state.
\begin{figure}
\epsfxsize 5cm
\epsfbox{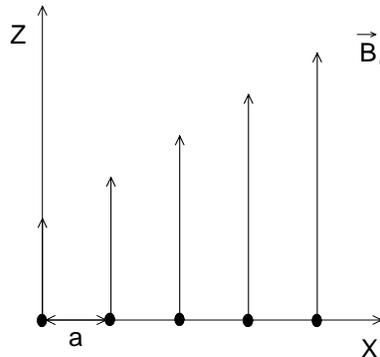}
\caption{Fig. 2:~The magnitude of the permanent magnetic field slightly 
increases in the $x$-direction. $a$ is the distance between the 
neighboring atoms.}
\label{fig.1}
\end{figure}

\section{A Control-Not gate}

The ability to implement a CN gate is important for quantum 
computation as any quantum algorithm can be decomposed into a 
sequence of one-qubit rotations and CN gates \cite{6}. We propose to 
use a scheme close to that outlined in our MRFM proposal \cite{1,2}. 
It consists of three steps:

1) One applies an electronic $\pi$-pulse with the frequency 
$f=f_{e1}(x_k)$. This pulse drives the electron spin of the ion ``k'' 
into the excited state only if the nuclear spin of the ion ``k'' (a 
control qubit) is in the excited state. The hyperfine splitting of 
the ion must be smaller than the difference, $\Delta f_e$, of 
precession frequencies of neighboring ions. For $\partial 
B_0/\partial x=10^5$T/m we have: $\Delta f_e=14$MHz. So, this 
condition is satisfied, and the frequency, $f_{e1}(x_k)$, is unique 
in the spin chain. (The dipole-dipole contribution to the oscillation 
is small in comparison to the hyperfine splitting.)

2) One tunes the frequency of the nuclear $\pi$-pulse for the ion 
``k+1'' (or ``k-1'') taking into consideration the dipole field 
produced by an electron spin on the nuclear spin of the ion ``k+1'' 
(a target qubit). If the electron spin of the ion ``k'' did not 
change its direction, the NMR frequency for the ion ``k+1'' (taking 
into account a dipole field of electron spins) is,
$$
f=f_n(x_{k+1})-f_{nd}-f^\prime_{nd},\eqno(8)
$$
where $f_{nd}$ is the dipole shift due to two neighboring electron 
spins, and $f^\prime_{nd}$ is the dipole shift due to all other 
electron spins. (For $a=5$nm, $f_{nd}\approx 200$Hz, 
$f^\prime_{nd}<40$Hz, the nuclear-nuclear dipole interaction is 
negligible.) If the electron spin of the ion ``k'' has changed its 
direction, then the NMR frequency for the ion ``k+1'' is,
$$
f=f_n(x_{k+1})-f^\prime_{nd}.\eqno(9)
$$
The difference, $f_{nd}$, between the frequencies (8) and (9) must be 
smaller than $\Delta f_n$. In our case this condition is satisfied, 
so both frequencies (8) and (9) are unique in a spin chain. If the 
frequency of a nuclear $\pi$-pulse is tuned to (9), and the Rabi 
frequency, $f_{nR}$, of the nuclear $\pi$-pulse is less than 
$f_{nd}$, then the nuclear spin of the ion ``k+1'' changes its 
direction only when the control qubit is in the excited state. This 
provides an implementation of a CN gate. This step imposes a most 
severe restriction on the duration of a quantum gate: $\tau=\pi/2\pi 
f_{nR}>2.5$ms. The same step imposes a restriction on the electron 
relaxation time which must be greater than $\tau$. (Available data 
\cite{7} show that the electron relaxation time for paramagnetic 
impurities can be much greater than $2.5$ms.)

3) Finally, one repeats the first step to drive an electron spin of 
the ion ``k'' back to the ground state, if the control qubit is in 
the exited state.

\section{Conclusion}

We proposed a nuclear spin solid-state quantum computer based on 
scanning tunneling microscopy (STM). Our proposal does not 
require sophisticated single-atom electrostatic gates. A single-spin 
measurement is provided by measuring the modulation of the tunneling 
current. Selective excitation of a nuclear spin can be achieved using 
an inhomogeneous permanent magnetic field. A Control-Not gate is 
provided by utilizing the dipole-dipole interaction between the 
electron and nuclear spins.
Our envisioned STM quantum computer would operate at high magnetic 
field ($\sim 10$T) and at low temperature $\sim 1$K.
\\ \ \\    
\quad\\
{\large\bf Acknowledgments}\\ \ \\
This work  was supported by the Department of Energy (DOE) under 
contract W-7405-ENG-36. The work of GPB and VIT was partly supported 
by the National Security Agency (NSA) and by the Advanced Research and 
Development Activity (ARDA).
\end{document}